\documentclass[12pt]{article}
\usepackage{epsf}
\usepackage{cite}

\def\ber{\begin{eqnarray}}
\def\eer{ \end{eqnarray}}  
\def\be{\begin{equation}}
\def\ee{\end{equation}}
\def\la{\label}

\def\lbr{\lbrack} 
\def\rbr{\rbrack} 

 \def\n{\nonumber\\}

\def\bigone{\hbox{1\kern -.23em {\rm l}}}     
\def\ZZ{\hbox{\zfont Z\kern-.4emZ}}

  \def\CF {{\cal F}}
  
  \def\CH {{\cal H}}

  \def\CL {{\cal L}}
  
  \def\CN {{\cal N}}

\catcode`@=11 \@addtoreset{equation}{section} \catcode`@=12

\begin{document}

\setlength \arraycolsep{2pt}

\begin{titlepage}
\vfill
\begin{center}
{\Large \bf Metallic and antiferromagnetic fixed points from gravity}\\[1cm] 
Chandrima Paul {\footnote{mail:plchandrima@gmail.com}} \\ 
\vskip3mm
\emph{ Department of Physics,\\ Sikkim University, 6th Mile, Gangtok 737102}
\end{center}
\vfill

\begin{abstract} 
We consider SU(2)$\times$ U(1) gauge theory coupled to matter field in adjoints and study RG group flow. We constructed Callan Symanzik equation and subsequent $\beta$ functions and study the fixed points. We find there are two fixed points, showing metallic and antiferromagnetic behaviour. We have shown that metallic phase develops an instability if certain parametric conditions are satisfied.
\end{abstract}
\vskip2mm
11.15.Ex Spontaneous breaking of gauge symmetries;
\\
11.25.Tq Gauge/string duality  
\vfill 
\end{titlepage}

\begin{center}
\tableofcontents
\end{center}
\newpage

\section{Introduction}
In last two decades there is significant development in the study of holographic renormalization group (\cite{kiritsis}-\cite{bianchi}).  In 1999 Verlinde et.al. proposed  the formalism \cite{lectholo2,verlinde,papa,bianchi} which has wide applications in AdS/CFT correspondence.   According to it an on shell action can be written as the sum of local and non local part.  Buik equations of motion can be written using Hamilton Jacobi formalism so that
they look formally similar to RG equations. Many other works also have appeared in this context \cite{notes1,notes2} (and references therein). Subsequently, studies of RG group  in the context of dilaton theory appeared in literature \cite{kiritsis,exotic,dyonic,bourdier}. It was developed futher to include the gauge fields\cite{kikuchi}.
\vskip1mm
\hskip2mm
AdS/CFT correspondence has been applied successfully in the realm of condensed matter physics \cite{Hartnoll:2009sz,Herzog:2009xv,Horowitz:2010gk,Hartnoll:2011fn,Sachdev:2011wg,Green:2013fqa} (and references therein), and especially in the context of holographic superconductors (\cite{buildingads}-\cite{pufu}). High temperature superconductors have a rich phase structure and it is interesting to understand transitions among the various phases from a holographic perspective. RG flow has been used in the study of such transitions. In particular, transition between metallic and insulating phase was analysed in \cite{metal2insulator}, where these two phases were described as fixed points of RG flow. Similar works in the study of metallic and insulaing fixed point also appeared in \cite{metalinsulator}.
\vskip1mm
\hskip2mm
In the present work, we will consider some holographic helical superconductor \cite{metal2insulator,gaun,t,lett}. In condensed matter theory there exist p-wave superconducting phases with a helical order. This phase in gravity is described by the black hole with helical symmetry.  More precisely, for five dimensional black hole, if $x^1, x^2,x^3$ describes the transverse directions, the killing vectors associated with the black hole are given by $\partial_{x^2},\partial_{x^3}$ and $\partial_{x^1}-p({x^2}\partial_{x^3} -{x^3}\partial_{x^2})$, where p is a constant(known as the pitch of the helix), which generates a helical motion consisting of a translation in the $x^1$ direction combined with a simultaneous rotation in $(x^2,x^3)$ plane. These systems break translational invariance with retaining homegeneity and so it may admit insulating phase. The connection of helical black holes with the insulating phase of high temperature superconductor is explored in \cite{metal2insulator}.
 \vskip1mm
\hskip2mm
As explained in \cite{iqbal} holographic superconductor  may admit antiferromagnetic phase through condensation of adjoint scalar by breaking  $SU(2)\rightarrow U(1)$. In this connection, we introduced  a model  \cite{oupar} with gravity coupled to $SU(2)\times U(1)$ gauge theory along with a scalar field in adjoint, which admits helical superconducting black hole. For certain field configuration, the near horizon geometry of the black hole solution of this model admits metallic phase\cite{fermi}.  Also it may admit insulating antiferromagnetic phase through condensation of adjoint scalar by breaking  $SU(2)\rightarrow U(1)$. In that model we would like to explore the possiblity of a phase transition between metallic and an insulating phase with antiferromagnetic order. In order to do that we have applied holographic RG flow following \cite{lectholo2,verlinde}.  According to holographic duality a stable phase corresponds to certain stable field configuration and appear as the fixed point of holographic RG group flow.  An instability  of the phase corresponds to flow between one  fixed point to the other, i.e one geometric configuration and vacuum field configuration to the other. We have applied the technique of holographic RG flow in the present model. We have obtained the potential and by minimising it we have identified the fixed points as metallic and antiferromagnetic phases. 
\vskip1mm
\hskip2mm
The organization of the present work is as follows. In the  section 2 we develop RG group formalism for our model, obtain the Callan-Symanzik equation.  In the  section 3 we develop the potential equations.  In section 4 we obtain metallic fixed point.  In  section 5 we show that the other fixed point has an antiferromagnetic behaviour.   In section 6, in a different analysis we show that there exist a condition,  when metallic phase develops an instability.  Finally we come to conclusion in section 7.   In the appendix, we apply our method to the equation of motion to obtain the behaviour of some of the fields and metric components near critical point in terms of the potentials.

\section{Derivation of  Callan Symmanzic equation }

We coinsider the action\cite{oupar},
\ber
S &=&  \int_M d^5 x \sqrt{-g} \lbrack R + 12-{\frac{1}{4}} F_{\mu \nu}F ^{\mu \nu}- {\frac{1}{4}}W_{\mu\nu}W^{\mu\nu}
- {\frac{1}{2}}(D_\mu \phi^a)^\dag (D_\mu \phi^a) - {\frac{m^2}{2}}\phi^{a\dag}\phi^a\rbrack\n
 &-&{\frac{\kappa}{2}}\int B\wedge F \wedge W +\int_{\partial M}\sqrt{-\gamma} d^4 x 2K .
\la{action}
\eer
where $F = dA, W^a = dB^a -\epsilon^{abc} B^b \wedge B^c$ and $(D_\mu \phi)^a = \partial_\mu \phi^a + iA_\mu\phi^a - \epsilon^{abc} B_\mu^b\phi^c$,   with A and B are U(1) and SU(2) gauge fields, $\phi$ is the scalar in SU(2)$\times$ U(1) adjoint representation. Last term is 
the Gibbons-Hawking term and $\gamma^{ij}$ is boundary metric.  
The metric ansatz 
\ber
ds^2 
&=& -f(r) dt^2 + {\frac{dr^2}{U(r)}} + e^{2 v_1 }dx_1^2 + \left(e^{2 v_2} Cos^2 px_1 + e^{2 v_3} Sin^2 px_1\right) dx_2^2 \n
&+& \left(e^{2 v_3} Cos^2 px_1 + e^{2 v_2} Sin^2 px_1\right) dx_3^2 + 2(e^{2v_3}-e^{2v_2})Cos( px_1 ) Sin( px_1) dx_2 dx_3,
\la{metric}
\eer
where $v_i(r),U(r)$ are radial functions,   define asymptotically AdS geometry, which implies at $r\rightarrow\infty$,  we have $v_i(r)\rightarrow ln r$ and $U(r)= r^2 $.   p is a parameter.
Furthermore, for our purpose we impose simplifying ansatz for the fields, which is consistent with the equations of motion, as given in the following:
\be
B^1 = w(r)\omega_2, \quad B^2 = B^3 = 0\quad,\quad
\phi^1=\phi^2=0,\quad \phi^3=\phi(r), \quad A = a(r)dt,
\la{ansatz}
\ee
where we use one forms $\omega_1 = dx^1$,  $\omega_2 = \cos(px^1)dx^2 - \sin(px^1)dx^3$, 
$\omega_3 =  \sin(px^1)dx^2+ \cos(px^1)dx^3$,

\vskip1mm
\hskip3mm
In order to formulate the bulk dynamics in a hamiltonian language we begin by decomposing
the bulk variables in components along and transverse to the radial coordinate,  as in the standard
ADM treatment of gravity,  except that the hamiltonian time now is the radial coordinate
instead of real time.   In particular the bulk metric is written in the form\cite{dyonic},\cite{lectholo2}  

\be
ds^2 = (N^2 + N^i N^i)dr^2 +2N^i drdx^i +\gamma_{ij}dx^i dx^j.
\la{admmetric}
\ee  
Clearly one can recast the above metric (\ref{metric}) in the above form (\ref{admmetric}) with a choice $N^i = 0$ which will be our gauge choice.
The inverse metric of (\ref{admmetric})   is given by 
\be
g^{rr} = {\frac{1}{N^2}}\,\,;\,\,\,\,\, g^{ri}= -{\frac{N^i}{N^2}}\,\,;\,\,\,\,\,g^{ij}=\gamma^{ij} +{\frac{N^i N^j}{N^2}}.
\la{inverseadmmetric}
\ee 
In terms of the above metric, the  Ricci scalar can be written as
\be
R(g)=R(\gamma)+K^2 - K_{ij}K^{ij}+\nabla_\mu(-2Kn^\mu +2 n^\nu \nabla_\nu n^\mu),
\la{k}
\ee
where $K_{ij}$ is extrinsic curvature given by
\be
K_{ij} = {\frac{1}{2N}}(\dot{\gamma}_{ij} - D_i N_j - D_j N_i),
\la{extrinsic curvature}
\ee
and $n^\mu = \left({\frac{1}{N}},-{\frac{N^i}{N}}\right)$is the unit normal vector to the constant r hypersurfaces. In the above, $D_i$ represents covariant derivative with respect to the induced metric $\gamma_{ij}$.
The reduced action from (\ref{action}) is given by

\ber
S &=&\int d^4 x N\sqrt{\gamma}\lbrack R(\gamma) + K^2 - K^{ij}K_{ij}-{\frac{1}{2N^2}}(D_r \phi- N^iD_i\phi)^\dag(D_r \phi- N^j D_j\phi)
\n
  &-& {\frac{m^2}{2}} \phi^\dag \phi -\gamma^{ij} (D_i \phi)^\dag D_j \phi\n
  &-& {\frac{1}{2}} {\frac{\gamma^{ij}}{N^2}}\left( F_{ri} -N_k F^k_i \right)\left( F_{rj} -N_m F^m_j \right)\n
  &-& {\frac{1}{2}} {\frac{\gamma^{ij}}{N^2}}\left( W_{ri} -N_k W^k_i \right)\left( W_{rj} -N_m W^m_j \right)\n
  &-& {\frac{1}{4}}F_{ij}F^{ij}-{\frac{1}{4}}W_{ij}W^{ij}  \rbrack \n
  &-& \kappa \epsilon^{rijkm}\left(\partial_r A_i - \partial_i A_r-N_k F^k_i \right)W_{jk}B_m .
 \la{reducedaction}
\eer
One can show that for on shell action S,  the canonically conjugate momenta p of  any field q is given by \cite{lectholo2,kiritsis}
\be
p={\frac{\partial S}{\partial q}}.
\la{onshellmomenta}
\ee
Now from(\ref{admmetric}), since, N, $N^i$ are cyclic, so
\be
H = N \CH + {\CN}^i {\CH}^i + a_r \CF = 0 \,\,\,\,\,\,\, \Rightarrow \CH=\CH^i=0,
\la{hamilton constraint}
\ee

where
\ber
\CH &=&- {\frac{1}{\sqrt{-\gamma}}}\left(\gamma_{ik}\gamma_{jl} -{\frac{1}{d-1}}\gamma_{ij} \gamma_{kl}\right)\pi^{ij}\pi^{kl}
   - 2\pi_\phi^{\dag a}\pi_\phi^{ a}-{\frac{1}{2}}\gamma_{ij}\pi_B^i \pi_B^j\n
    &-&{\frac{1}{2}}\gamma_{ij}\left(\pi^i_A + {\frac{\kappa}{2}}\epsilon^{rinkm} W_{nk} B_m \right)\left(\pi^j_A + {\frac{\kappa}{2}}\epsilon^{ri{n^\prime}{k^\prime}{m^\prime}} W_{{n^\prime}{k^\prime}} B_{m^\prime} \right)\n
    &+&{\sqrt{-\gamma}}\lbrace \gamma^{ij}(D_i \phi^a)^\dag (D_j \phi^a) + R + 12 \n
    &+& {\frac{1}{4}}F_{ij}F^{ij} + {\frac{1}{4}} W_{ij}^a W^{a\,\,ij} + {\frac{m^2}{2}}\phi^{a\dag}\phi^a\rbrace. 
\la{curlh}
\eer
\ber
\CH^i&=& D_j\pi^{ij}+(\pi_\phi^\dag D^i\phi^a + \pi_\phi^a D^i \phi^\dag)
     +F^{ij}(\pi_A^s+{\frac{\kappa}{2}}\epsilon^{rsnkm} W_{nk}B_m)\gamma_{sj}    \n
     &+& {\frac{1}{2}}W^{ij,a}\pi_{B,m}^a\gamma^{jm},
\la{hi}
\eer
where $\pi^{ij}$ correspond to canonically conjugate momentum of the metric $\gamma^{ij}$,  $\pi_B,  \pi_A$ correspond to canonical conjugate momentum of SU(2) gauge field B and U(1) gauge field A and $\pi_\phi$ is the canonically conjugate momentum to $\phi$.
 Finally $\CF$ is the coefficient of radial component of U(1) gauge field in the expression of hamiltonian and it vanishes, since we consider U(1) gauge field in time direction only.
We have 
\be
\CH = \lbrace S,S\rbrace -  \CL_d,
\la{lagrangerelation}
\ee

where $\CL_d$ is the d dimensional part of the lagrangian (d corresponds to boundary dimension) and the bracket expressed as:
\ber
\left\lbrace S, S\right\rbrace
&=& - {\frac{1}{\sqrt{-\gamma}}}\lbrack\left( 2{\gamma_{i k}}{\gamma_{j l}}- {\frac{1}{d-1}}\gamma_{ij} \gamma_{k l}\right)\pi^{i j} \pi^{k l} - 2\pi_{\phi^\dag}^a \pi_\phi^a - {\frac{1}{2}}\gamma_{ij} \pi_B^i\pi_B^j\n
&-& {\frac{1}{2}} \gamma^{ij} \left( \pi_i^A + {\frac{\kappa}{2}}\epsilon^{rinkm} W_{nk} B_m \right) \left( \pi_j^A + 
\epsilon^{r j n^\prime k^\prime m^\prime}W_{n^\prime k^\prime}B_{m^\prime} \right)\rbrack.
\la{bracketrelation}
\eer

$\CH = 0$ gives

\be
\left\lbrace S,S\right\rbrace = \CL_d.
\la{bracketrelat}
\ee
Now as proposed by Verlinde \cite{verlinde}, the on shell action  can be written as a local and non local part on boundry, i.e
\be
S= S_{loc} + \Gamma,
\la{localnonlocal}
\ee
where $S_{loc}$ is the local part given by

\ber
S_{loc}
&=& S_{loc}^{(0)} + S_{loc}^{(2)}+.............\n
S_{loc}^{(0)}
&=&\int \sqrt{-\gamma}W\left(\phi^\dag (r) \phi(r)\right)\n
 S_{loc}^{(2)}
&=&\int \sqrt{-\gamma}\left\lbrace \Phi\left(\phi^\dag (r) \phi(r)\right) R_d + M\left(\phi^\dag (r) \phi(r)\right) \left((D_i\phi (r))^\dag (D_i\phi(r))\right)\right\rbrace,
\la{localaction}
\eer
where in the above the superscripts denote the number of derivatives/weights and $R_d$ is d dimensional Ricci scalar.  We consider most general expansion of $S_{loc}$ as considered (\cite{kikuchi}, \cite{verlinde}).  Also $\Gamma$ is the quantum effective action. For d dimensional lagrangian.  We  write
\ber
\CL_d^{(0)}
&=&\left\lbrace d(d-1) -{\frac{m^2}{2}} \left(\phi^\dag (r) \phi(r)\right)\right\rbrace\n
 \CL_d^{(2)}
&=&\left\lbrace R_d + \gamma^{ij} \left((D_i \phi (r))^\dag (D_j \phi(r))\right)\right\rbrace\n
\CL_d^{(4)}
&=& -{\frac{1}{4}} F_{ij}F ^{ij}- {\frac{1}{4}}W_{ij}W^{ij},
\la{derivativeaction}
\eer
where $D_i \phi (r)$ is d dimensional covariant derivative of $\phi$ and according to our ansatz(\ref{metric},\ref{ansatz}),  is  given by
\be
(D_i\phi)^a(D_i\phi)^a = \left(w(r)^2  e^{-2 v_2}  -{\frac{a^2 (r)}{f(r)}}\right)\phi^\dag (r) \phi(r).   
\la{dcovariantderivative}
\ee
So the Hamilton -Jacobi equation reduces to the equations (\ref{bracketrelat},\ref{localaction},\ref{derivativeaction})
\ber
\left\lbrace S_{loc}^{(0)},S_{loc}^{(0)}\right\rbrace 
&=& \CL_d^{(0)}\\
2 \left\lbrace S_{loc}^{(0)},S_{loc}^{(2)}\right\rbrace
&=& \CL_d^{(2)}\\
\left\lbrace S_{loc}^{(2)},S_{loc}^{(2)}\right\rbrace + 2 \left\lbrace S_{loc}^{(2)}, \Gamma\right\rbrace
&=& \CL_d^{(4)}.
\la{workingprinciple}
\eer
The radial derivative of the fields are given by formal hamiltonian equation $${\frac{d\phi}{dr}} = {\frac{\partial H}{\partial \pi_\phi}}={\frac{\partial H}{\partial ({\frac{\partial S}{\partial \phi^\dag }})}}$$
where by S we refer to on Shell action. So inserting the expansions(\ref{localaction}), we can write (\cite{dyonic},\cite{kikuchi},\cite{lectholo2}) 

\ber
{\frac{d\phi}{dr}}  &=&{\frac{\partial W}{\partial \phi}}\n
{\frac{d w(r)}{dr}} &=& {\frac{\partial (M(\phi^\dag \phi)(D_i \phi)^\dag(D^i \phi)}{\partial w(r)}}=M(\phi^\dag \phi)\,Exp(-2v_2(r))\,w(r)\,\phi^\dag \phi\n
{\frac{d a(r)}{dr}} &=& {\frac{\partial (M(\phi^\dag \phi)(D_i \phi)^\dag(D^i \phi)}{\partial a(r)}}=-M(\phi^\dag \phi)\,{\frac{1}{f(r)}}\,a(r)\,{\overline{\phi}}\phi,
\la{fieldderivative}
\eer
and finally for the gravity part it follows from \cite{lectholo2} that
\ber
{\frac{\partial{g_{\mu\nu}}}{\partial r}}g^{\mu\nu}
&=& (-2 {\pi_{\mu \nu}} + {\frac{2}{d-1}}{\pi^\lambda_\lambda}g_{\mu\nu})g^{\mu\nu}\n
&=& {\frac{2}{d-1}}{\pi^\lambda_\lambda}\n
&=& g^{\mu\nu} {\frac{2}{d-1}}{\frac{\partial}{\partial {g_{\mu\nu}}}}\lbrace\int\sqrt{g}W(\phi)\rbrace \n
\Rightarrow {\frac{ d g_{\mu \nu}}{dr}}
&=& -{\frac{W({\overline{\phi}}\phi)}{d-1}}{g_{\mu \nu}}+ C(r),
\la{gravityequation}
\eer

where one can introduce the constant $C(r)$, which is independent  of $g_{\mu\nu}$, but, it can differ for different component of metric, actually gives anisotropy of the metric around the fixed point.  However for simplicity of our solution we will consider $C = 0$ for all metric.
\\
It is also important to note that, since the field derivatives are obtained from $W(\phi^\dag \phi),M(\phi^\dag \phi)$ in 
(\ref{fieldderivative},\ref{gravityequation}), we will effectively treat these terms as potential and call them by potential in the rest of the article. 
Now, it implies, from (\ref{gravityequation})
\be
{\frac{U^\prime}{U}}= {\frac{W(\phi^\dag \phi)}{d-1}}\,\,\,\,;\,\,\,\,v_i^\prime = - {\frac{W(\phi^\dag \phi)}{2(d-1)}}\,\,\,\,;\,\,\,\,{\frac{f^\prime}{f}}= -{\frac{W(\phi^\dag \phi)}{d-1}}.
\la{metricrelation}
\ee

Now, in order to write generalized expression for the bracket, following \cite{kikuchi}, \cite{dyonic} we will assign weight to all fields.  This is as following:
\ber
\gamma_{\mu \nu},\phi,\Gamma
&   &: 0  \n
\partial_\mu, B_\mu , A_\mu 
&   &: 1\n 
R, R_{\mu \nu},R_{\mu \nu \rho \sigma}, \partial^2
&   &: 2\n
{\frac{\partial}{\partial a(r)}},{\frac{\partial}{\partial w(r)}}
&   &: d-1\n
{\frac{\partial}{\partial h_{\mu\nu} (r)}},{\frac{\partial}{\partial \phi(r)}}
&   &: d.
\la{weight}
\eer
\be  
S_{loc;{w-d}} \,\,\,=\,\,\, \int d^d x \sqrt{-\gamma}L^{(w)}.
\la{wd}
\ee

Finally we write the expression for generalized bracket as
\be
2\left\lbrace S_{loc}, \Gamma \right\rbrace_d = -\left\lbrace S_{loc},S_{loc} \right\rbrace_d +\CL_d^{(d)},
\la{cseqnrelation}
\ee
where $\left\lbrace S_{loc}, \Gamma \right\rbrace_d$ and $\left\lbrace S_{loc},S_{loc} \right\rbrace_d $ denotes weight d term from the bracket.  Also $\CL_d^{(d)}$ term denotes weight d term of $\CL_d$. 
Finally we write (\ref{cseqnrelation}) as generalized bracket expression

\ber
& & {\frac{W\left(\phi^\dag(r)\phi(r)\right)}{d-1}} \gamma^{kl} {\frac{\partial \Gamma}{\partial \gamma^{kl}}} - {\frac{\partial W\left(\phi^\dag(r)\phi(r)\right) }{\partial \phi^{\dag}}}{\frac{\partial \Gamma}{\partial \phi}} -{\frac{\partial W\left(\phi^\dag(r)\phi(r)\right) }{\partial \phi}}{\frac{\partial \Gamma}{\partial \phi^\dag}} \n
&-& \gamma^{ij}{\frac{\partial S_{loc;2-d}}{\partial B^i}}{\frac{\partial \Gamma}{\partial B^j}}-\gamma^{ij}{\frac{\partial S_{loc;2-d}}{\partial A^i}}{\frac{\partial \Gamma}{\partial A^j}}\n
&=& {\frac{1}{2}} (\CL^{(d)}_d)-{\frac{1}{2}}\left\lbrace S,S\right\rbrace_d +{\frac{1}{2}}\kappa\epsilon^{rijkm}W_{jk}B_m {\frac{\partial S_{loc}}{\partial A^i}}.
\la{cseqn}
\eer

We write Callan-Symmanzik equation after some rearrangement
\ber
& & \gamma^{kl} {\frac{\partial \Gamma}{\partial \gamma^{kl}}} - \beta_\phi {\frac{\partial \Gamma}{\partial \phi}} -{\beta_{\phi^\dag}} {\frac{\partial \Gamma}{\partial \phi^\dag}} 
- \beta_B {\frac{\partial \Gamma}{\partial B^j}}-\beta_{a(r)} {\frac{\partial \Gamma}{\partial A^j}}\n
&=& {\frac{1}{2}}\left\lbrace{\frac{W\left(\phi^\dag(r)\phi(r)\right)}{d-1}} \right\rbrace ^{-1} \left\lbrack (\CL^{(d)}_d)-\left\lbrace S,S\right\rbrace_d +\kappa\epsilon^{rijkm}W_{jk}B_m {\frac{\partial S_{loc}}{\partial A^i}}\right\rbrack, 
\la{cseqnbeta}
\eer

with
\ber
\beta_\phi 
&=& \left\lbrace{\frac{W\left(\phi^\dag(r)\phi(r)\right)}{d-1}} \right\rbrace^{-1} {\frac{\partial W\left(\phi^\dag(r)\phi(r)\right) }{\partial \phi^{\dag}}}\n
\beta_{\phi^\dag}
&=&\left\lbrace{\frac{W\left(\phi^\dag(r)\phi(r)\right)}{d-1}} \right\rbrace^{-1} {\frac{\partial W\left(\phi^\dag(r)\phi(r)\right) }{\partial \phi}} \n
\beta_{a(r)}
&=&\left\lbrace{\frac{W\left(\phi^\dag(r)\phi(r)\right)}{d-1}} \right\rbrace^{-1} \gamma^{ij}{\frac{\partial S_{loc;2-d}}{\partial A^i}} \n
\beta_{w(r)}
&=&\left\lbrace{\frac{W\left(\phi^\dag(r)\phi(r)\right)}{d-1}} \right\rbrace^{-1} \gamma^{ij}{\frac{\partial S_{loc;2-d}}{\partial B^i}}. 
\la{betafunction}
\eer

Now the fixed points corresponds to simultaneous zero of the $\beta$ functions.  Clearly $\beta_\phi = \beta_{\phi^\dag}=0$ implies, the   potential $W\left(\phi^\dag(r)\phi(r)\right)$ is at its extrema. To understand it for SU(2) gauge field and vector field, recall (\ref{wd},\ref{localaction});  while $\beta_{w(r)} = 0 $, can be achieved by setting $w(r)=0$, however the vanishing of $\beta_{a(r)}$ needs, for nonzero $\phi$,
\be
M\left(\phi^\dag(r)\phi(r)\right)=0
\la{mzero}
\ee
at fixed point.
We will elaborate these points further in section 4 and 5.

\section{The potential equations}
The potential equations, i.e the equations for $W(\phi^\dag \phi),M(\phi^\dag \phi)$  in (\ref{fieldderivative},\ref{gravityequation}), are direct consequence of the equation $\CH=0$ (\ref{bracketrelat}), where the canonically conjugate momenta is expressed as the derivative of on shell action(\ref{onshellmomenta}), and the on shell action is expanded as (\ref{localaction}).  The potential equation at zeroth derivative order is expressed as 
\be
{\frac{d}{4(d-1)}}\lbr W({\phi^\dag} \phi)\rbr^2 -{\frac{\partial W({\phi^\dag} \phi)}{\partial \phi^\dag}}{\frac{\partial W({\phi^\dag} \phi)}{\partial \phi}} -{\frac{1}{2}}{\phi^\dag}{\phi}\,M({\phi^\dag} \phi)W({\phi^\dag} \phi)\left(e^{-2{v_2}}-{\frac{1}{f(r)}}\right) \,\,= \,\,V({\phi^\dag}\phi).   
\la{zeroderivative}
\ee
At two derivative level the expression is
\ber
-{\frac{1}{2}}({D_i}\phi)^\dag (D^i \phi) + R_d
&=&{\frac{d-2}{d-1}}\left\lbrack \Phi(\phi)\,{R_d}+M(\phi^\dag \phi)(D_i \phi)^\dag(D^i \phi)\right\rbrack W(\phi^\dag \phi)
-\left\lbrace{\frac{\partial W(\phi^\dag \phi)}{\partial \phi^\dag}}{\frac{\partial \Phi(\phi^\dag \phi)}{\partial \phi}}\right\rbrace R_d\n
&-&\left\lbrace{\frac{\partial W(\phi^\dag \phi)}{\partial \phi^\dag}}{\frac{\partial M(\phi^\dag \phi)}{\partial \phi}}\right\rbrace ({D_i}\phi^\dag)({D^i}\phi)\n
&+& M(\phi^\dag \phi){\frac{\partial W(\phi^\dag \phi)}{\partial \phi^{a\,\dag}}}\left\lbrace -{\frac{a(r)}{f(r)}}(D_i \phi)^a + f^{bca}(D_i \phi)^b {B^c}(r)\right\rbrace \n
&-& W(\phi^\dag \phi) M(\phi^\dag \phi)\left\lbrace -{\frac{a(r)^2}{ f(r)}} + e^{-2 {v_2}}w(r)^2 \right\rbrace\n
&-&{\frac{1}{2}} M({\phi^\dag \phi}) e^{-4{v_2}(r)}{[\phi^\dag \phi]}^4 ]\n
&-& {\frac{1}{2}}\lbrace \left(-{\frac{1}{f(r)}}{(\phi^\dag \phi)}^2 M({\phi^\dag \phi}) + {\frac{\kappa}{2}}\epsilon^{r i n k m}(W_{nk} b_m)\right)  (-{\frac{1}{f(r)}}{(\phi^\dag \phi)}^2 M({\phi^\dag \phi}) \n
&+& {\frac{\kappa}{2}}\epsilon^{r i {n^\prime}{ k^\prime} {m^\prime}}(W_{{n^\prime}{k^\prime}} b_{m^\prime}) ) \rbrace.  
\la{twoderivative}
\eer

\section{Critical Points : AdS fixed point}
Critical points are given by simultaneous zero of $\beta$ functions.  From (\ref{betafunction}), first condition for zero of the $\beta$ function is extrema of $W(\phi^\dag \phi)$.  In order to find the critical points,  we need to solve the potential equation (\ref{zeroderivative})order by order.  
We expand
\be
W(\phi^\dag \phi)=W_o + W_1 (\phi^\dag \phi) + W_2 (\phi^\dag\phi)^2 +.....,
\la{expansion}
\ee
where $W_o,W_1,W_2$ are constant coefficients.
Clearly extremizing the above(\ref{expansion}), w.r.t. $\phi$, one trivial fixed point can be found at $\phi=0$.  Combining,(\ref{localaction}), (\ref{dcovariantderivative}),  (\ref{wd}),  (\ref{betafunction}),  (\ref{expansion}), we find $\phi=0$ is also the zero of the $\beta$ function associated with SU(2) gauge field and U(1) gauge field.   Moreover we can choose, at this fixed point SU(2) gauge field $w(r)=0$, then from  the action (\ref{action}) it is evident that this has RN AdS black hole solution \cite{fermi}.   Now according to \cite{fermi}, AdS RN black hole has a near horizon geometry given by $AdS_2 \times R^3$ which is dual to the metallic phase.   So we see that our first fixed point does have correspondence to metallic phase. 
\\
Now we solve the potential equation (\ref{zeroderivative}) order by order.   

It is evident that first order term is 
\be
W_o = 2(d-1).
\la{firstorder}
\ee

Before proceeding to next order term, we consider the term 
${\frac{1}{2}}{\phi^\dag}{\phi}\,M({\phi^\dag} \phi)W({\phi^\dag} \phi)\left(e^{-2{v_2}}-{\frac{1}{f(r)}}\right)$ from (\ref{zeroderivative}).
Near AdS background $e^{-2{v_2}}={\frac{1}{f(r)}}={\frac{1}{r^2}}$ (We consider AdS radius $R=1$). So exactly near AdS fixed point this term will not contribute.   We have the coefficient equation first order in $\phi^\dag\phi$ is
\be
{\frac{d}{4(d-1)}}(2W_o W_1) -2W_1^2=-{\frac{1}{2}}m^2 .
\la{firstordereqn}
\ee
We write
\be
W_1={\frac{1}{2}}\triangle_\pm .
\la{triangle}
\ee

Substituting (\ref{triangle}) in  (\ref{firstordereqn}) gives

\be
{\frac{d}{2}}\triangle_\pm -{\frac{1}{2}}\triangle_\pm^2=-{\frac{1}{2}}m^2 .
\la{adsrelation}
\ee

Near AdS fixed point,  we can ignore the quadratic and higher order terms in $\phi$ , so that we can write the potential $W(\phi^\dag \phi)$ as
\be
W(\phi^\dag \phi) = 2(d-1) + {\frac{1}{2}}\triangle_\pm \phi^\dag \phi,
\la{potentialexpression}
\ee
where $\triangle_\pm$ denotes the dimension of dual operator.
Near AdS fixed point the scalar field expression can be obtained by integrating $\phi$ derivative part of (\ref{fieldderivative}) 
\be
\phi(r)= A_1 r^{\triangle_+} + A_2 r^{\triangle_-}\,\,;\triangle_\pm = {\frac{d\pm \sqrt{d^2 +4m^2}}{2}},
\la{phiexpression}
\ee
To find an approximate expression of  $\omega(r),a(r)$ near fixed point, we follow (\ref{fieldderivative}),
\ber
{\frac{d \omega(r)}{dr}} &=& M(\phi^\dag \phi)\,Exp(-2v_2(r))\,w(r)\,\phi^\dag \phi= {\frac{{c_1}}{r^2}}({\phi^\dag \phi} - \alpha)(\phi^\dag \phi)\omega(r)\n
{\frac{d a(r)}{dr}} &=& -M(\phi^\dag \phi)\,{\frac{1}{f(r)}}\,a(r)\,\phi^\dag \phi={\frac{{c_1}}{r^2}}({\phi^\dag \phi} - \alpha)(\phi^\dag \phi)a(r).
\la{fieldderivativenew}
\eer

In the above we have substituted Ads background metric expression
$$Exp(-2v_2(r))={\frac{1}{f(r)}}={\frac{{1}}{r^2}}$$
Also we use $M(\phi^\dag \phi) = \phi^\dag \phi -\alpha$, which actually we have established in the next section (\ref{mexpansion}).
Integrating (\ref{fieldderivativenew}) we obtain series expression  
\ber
ln [\omega(r)]&=&-ln[a(r)]= {\frac{c_1}{r}}\lbrace \alpha(A_1 + A_2)^2 -(A_1 + A_2)^4\rbrace\n
              &+& c_1 ln r\lbrace -2\alpha (A_1 + A_2)(A_1 {\frac{\triangle_+}{2}} + A_2 {\frac{\triangle_-}{2}}) + 3(A_1 + A_2)^3 (A_1 {\frac{\triangle_+}{2}} + A_2 {\frac{\triangle_-}{2}})\rbrace\n
              &+& .......,
\la{omegaexpansion}
\eer
where $A_1, A_2$ are integration constant.   Similar expression can be derived for a(r).

\section{Nontrivial fixed point}
We are going  to show at nontrivial fixed point, the vacuum lies on a circle $\phi^\dag \phi = \alpha$, which implies $\phi= e^{i\chi(x)},  \phi^\dag = e^{-i\chi(x)}$,  where $\chi$ is any function of space-time variable x. Clearly this spontaneously breaks $SU(2)\times U(1)$ symmetry of the theory to $U(1)\times U(1)$.   Now following \cite{iqbal} this breakdown of symmetry $SU(2)\rightarrow U(1)$ with nontrivial v.e.v of the scalar,  in a model gravity coupled to SU(2) and U(1) gauge field with zero v.e.v of SU(2) gauge field,   gives rise to antiferromagnetism. 
We mentioned earlier (\ref{mzero}),  that  when the scalar field $\phi$ takes a nontrivial v.e.v, the zero of the $\beta$ function requires, the potential at two derivative order $M(\phi^\dag\phi) = 0$ . This requires $M(\phi^\dag\phi)$, to have a series form
\be
M(\phi^\dag\phi) = (\phi^\dag\phi - \alpha)+ O((\phi^\dag\phi - \alpha)^2) + ...,
\la{mexpansion}
\ee
where $\alpha$ is the value of v.e.v of $\langle \phi^\dag \phi \rangle$ at the nontrivial fixed point.   Also at this fixed point we have
$e^{-2v_2}\ne{\frac{1}{f(r)}}\ne {\frac{1}{r^2}}$.
We denote at the fixed point
\be
e^{-2v_2}-{\frac{1}{f(r)}}=c + O(\phi^\dag \phi)+....
\la{c}
\ee 

As before we can expand
$$W(\phi^\dag\phi)= W_o +{\frac{1}{2}} \triangle_\pm \phi^\dag \phi + c_1 (\phi^\dag\phi)^2 +...$$
Substitututing this in (\ref{zeroderivative}) and also using (\ref{mexpansion},\ref{c}) we obtain  the zeroth term $W_o = 2(d-1)$ and the term of the order $\phi^\dag\phi$ is given by
\be
d\triangle_\pm - \triangle_\pm^2 = -m^2 -2c\alpha(d-1).
\la{kitkit}
\ee
We see in the above expression that once we set $c=0$, we obtain the AdS RN expression (\ref{adsrelation}) which also gives AdS RN geometry (\ref {c}) near critical point.
 In order to find the expression of the extrema $\alpha$ we proceed in the following way:
\\
Extremizing $W(\phi^\dag\phi)$, we find,  one extrema lies at $\phi^\dag = 0$ (which corresponds to AdS fixed point) and the other one corresponds to
\be
{\frac{1}{2}}\triangle_\pm + 2c_1 \phi^\dag\phi=0
\la{nontrivialextreema}
\ee
Comparing (\ref{nontrivialextreema})with (\ref{mexpansion}), we obtain
\be
\alpha=-{\frac{\triangle_\pm}{4c_1}}.
\la{extreema}
\ee

So we have the equation for $\triangle_\pm$ is
\be
\triangle_\pm^2 +\left\lbr{\frac{c}{2c_1}}(d-1)-d\right\rbr\triangle_\pm -m^2 =0.
\la{t riangleeqn}
\ee

\ber
\triangle_\pm &=&-{\frac{b}{2}}\pm {\frac{1}{2}}\sqrt{b^2 +4m^2}\n
            b &=& \left\lbr{\frac{c}{2c_1}}(d-1)-d\right\rbr.
\la{b}
\eer

\section{Instability of metallic phase}
\vskip2mm
In  this section we are going to show that the metallic phase develops instability for certain parameter range and consequently develops  tendency to flow towards the other phase as we mentioned in the previous section, given by antiferromagnetic fixed point.   This is analogous to the theory of a high temperature superconductor, where the metallic phase develops instabilty at critical point and flow to insulating antiferromagnetic phase.
\\
We have mentioned that the metallic state at zero temperature is described by $AdS_2 \times R^3$ geometry, which is the near horizon geometry of AdS RN black hole as given by the AdS fixed point.  We consider deformation around this background in the following way\cite{metal2insulator},
\ber
U&=& 12r^2(1+ u_0 r^\delta)\,\,\,;\,\,\,v_i= v_o(1+v_{i1}r^\delta)\,\,\,;\,\,\,a= 2\sqrt{6}r(1+a_0 r^\delta)\n
w&=&w_o r^\delta\,\,\,;\,\,\,\phi=\phi_o r^\delta\,\,;\,\,f = 12r^2(1+ f_0 r^\delta)
\la{horizonperturbation}
\eer
The exponents are determined by the equations of motion and come in pairs corresponding to the power of two roots of the equations.
We find two marginal operator with $\delta_\pm = 0$ which corresponds to rescaling of $x_1, x_2, x_3$.  

The four remaining operator are given by positive weight
\ber
\delta_+
&=& 1\,\,\,\,;\,\,\,\,-{\frac{1}{2}}+\sqrt{{\frac{1}{4}} +{\frac{1}{12}}p^2 e^{-2v_0} -{\frac{\kappa}{\sqrt{6}}}pe^{-2v_0}}\,\,\,\,;\n
& &-{\frac{1}{2}}+\sqrt{{\frac{1}{12}}(1+m^2)}\,\,\,\,;\,\,\,\,{\frac{1}{2}}+\sqrt{{\frac{1}{4}}+{\frac{1}{3}}p^2 e^{-2v_0}}. 
\la{weight}
\eer
if
\be
2\sqrt{6}\kappa<pe^{-v_0}\,\,;\,\,m^2>2 .
\la{condition}
\ee
We see that all four modes are irrelevant with $\delta_+ >0$. We see that when both the  inequalities in (\ref{condition}) are satisfied,
the solution is translation invariant and stable.   The violation of second inequality make the respective deformation mode relevant and hence AdS RN solution unstable.  However the translation symmetry will remain unbroken because we have $w\rightarrow 0$ as $r \rightarrow 0$ and the respective deformation which causes the helical structure is being turned off.   So, in principle we can choose m so that the second inequality(involving m) is satisfied as otherwise violation of that inequality is not giving any interesting physics. The IR irrelevance of translation invariance breaking operators implies that phases with these interior geometries will be metals. The violation of the first inequality in (\ref{condition}) will cause $\delta_w <0$ about IR fixed point, make the operator relevant. When the deformation associated with $w(r)$ become relevant,w will be nonvanishing at horizon.  Consequently the translational symmmetry will break down and this instability mode will turn on the helical geumetry. So the IR  geometry will change from  $AdS_2 \times R^3$ to the one with helical symmetry.   Hence once this condition is violated, following \cite{metal2insulator}, we understand that system will flow away from metallic phase to insulating phase which will also be antiferromagnetic according to our identification.

\section{Conclusion}

Here we have developed RG group formalism for our proposed model SU(2)$\times$ U(1) gauge theory coupled to matter in adjoint representation.   We obtained the Callan Symanzik equation, beta functions.  We found the critical points, which corresponds to the   zero of the $\beta$ functions, corresponds to the extrema of the  potential W. 
 We observed the trivial fixed point corresponds to AdS RN black hole where the near horizon geometry corresponds to metallic phase. We also observed that our model has a nontrivial fixed point which breaks SU(2)$\times$ U(1) gauge symmetry to U(1)$\times$ U(1) symmetry and resembles antiferromagnetism. We have also checked that the metallic phase develops instability for certain parametric condition, We further want to comment that the study of thermodynamics at both the fixed point geometry will clarify the picture further.  We hope to report on this issue in future.   

\vskip6mm
\textbf{Acknowedgement}
\vskip1mm
It is a pleasure to thank Subir Mukhopadhyay for many useful discussions, valuable suggestions and encouragement. Author is also very much grateful to  Jerome P Gauntlett for valuble discussions and many useful suggestions.   Thanks to the organizers of the conference "Strings Gravity and Cosmology", Sep 2017, in Kyoto University, Japan, where the poster of this work was presented.
\vskip12mm
\appendix{\noindent {}}
\section{Appendix}
\setcounter{equation}{0}
\subsection{Equations of motion}

\ber
0 &=& a^{\prime \prime}(r)+\left\lbrace v_1^\prime + v_2^\prime + v_3^\prime +{\frac{1}{2}}\left({\frac{U^\prime}{U}}\,- {\frac{f^\prime}{f}} \right) \right\rbrace a^\prime (r) + e^{-v_1 (r) - v_2 (r) -v_3 (r)} \sqrt{{\frac{f}{U}}}  w(r) {w^\prime} (r) - \left\lbrace{\frac{\phi^\dag (r) \phi(r)}{U(r)}}\right\rbrace\,a(r),\n
0 &=&   w^{\prime \prime}(r)+\left\lbrace v_1^\prime + v_2^\prime - v_3^\prime +{\frac{1}{2}}\left({\frac{U^\prime}{U}}\,+ {\frac{f^\prime}{f}} \right) \right\rbrace w^\prime (r)           -p^2 e^{2(v_2-v_1-v_3)}{\frac{w(r)}{U(r)}}\n
  & & +p\kappa a'(r){\frac{w(r)}{U(r)}}e^{-(v_1-v_2-v_3)}-\phi^\dag\phi{\frac{w(r)}{U(r)}},\n
0 &=&\phi^{\prime \prime}(r)                      +\left\lbrace v_1^\prime + v_2^\prime + v_3^\prime +{\frac{1}{2}}\left({\frac{U^\prime}{U}}\,+ {\frac{f^\prime}{f}} \right) \right\rbrace \phi^\prime (r)
                                             +{\frac{a(r)^2}{U(r)^2}}\phi(r)-m^2 \phi(r)-{\frac{ w(r)^2}{U(r)}}e^{-2v_3 (r)}\phi(r),\n
0 &=&v_1^{\prime\prime}+v_2^{\prime\prime} + v_1^{\prime 2}+v_2^{\prime 2}+v_1^\prime v_2^\prime+\left({\frac{1}{2}}{\frac{U^\prime}{U}}+{\frac{1}{2}}{\frac{f^\prime}{f}}\right)(v_1^\prime+v_2^\prime)
    -{\frac{p^2}{2U}}e^{-2v_1}-{\frac{p^2}{4U}}e^{2(v_2-v_1-v_3)}+
{\frac{3p^2}{4U}}e^{2(v_2-v_1-v_3)}\n
  & & + {\frac{1}{2}}\left({\frac{f^{\prime\prime}(r)}{f(r)}} + {\frac{1}{2}}\left\lbrack{\frac{U^\prime(r)f^\prime(r)}{U(r)f(r)}}                        -{\frac{(f^\prime(r))^2}{(f(r))^2}}\right\rbrack\right)+{\frac{6}{U}}
    +{\frac{3}{8U}}{p^2}e^{-2v_1} [w_1(r)]^2 e^{-2v_3}         
                                             - {\frac{1}{8}}e^{-2v_2}[\partial_r w_1(r)]^2\n
  & & - {\frac{1}{4U}}w_1(r)\phi^\dag\phi e^{-2v_2}-{\frac{1}{4U}}{\partial_r\phi^\dag \partial_r\phi}
    +{\frac{[a(r)]^2}{4U(r) f(r)}}\phi^\dag\phi
+{\frac{1}{8}}a^{\prime}(r)^2 {\frac{1}{f(r)}} 
                                             +{\frac{1}{2 U(r)}}m^2 \phi^\dag \phi ,\n
0 &=&  v_1^{\prime\prime}+v_3^{\prime\prime}        + v_1^{\prime 2}+v_3^{\prime 2}+v_1^\prime v_3^\prime+\left({\frac{1}{2}}{\frac{U^\prime}{U}}+{\frac{1}{2}}{\frac{f^\prime}{f}}\right)(v_1^\prime+v_3^\prime)-{\frac{ p^2}{2U}}e^{-2v_1}\n
  & & +{\frac{3p^2}{4U}}e^{2(v_2-v_1-v_3)}-{\frac{p^2}{4U}}e^{2(v_3-v_1-v_2)}           + {\frac{1}{2}}\left({\frac{f^{\prime\prime}(r)}{f(r)}} + {\frac{1}{2}}\left\lbrack{\frac{U^\prime(r)f^\prime(r)}{U(r)f(r)}}                       -{\frac{(f^\prime(r))^2}{(f(r))^2}}\right\rbrack\right)+{\frac{6}{U}}\n
  & & +{\frac{3}{8}}e^{-2v_2} {w^{\prime 2}}
-{\frac{p^2}{8U}}e^{-2{(v_1 +v_3)}}w_1^2     +{\frac{1}{4U}}{\frac{a(r)^2}{f(r)}}\phi^\dag\phi-{\frac{1}{4}}{\partial_r\phi^\dag}{\partial_r\phi}+{\frac{1}{4U}}e^{-2 v_2}{w(r)^2}\phi^\dag \phi-{\frac{1}{4}}m^2 \phi^\dag\phi,\n
0 &=& v_2^{\prime\prime}+v_3^{\prime\prime}        + v_2^{\prime 2}+v_3^{\prime 2}+v_2^\prime v_3^\prime+\left({\frac{1}{2}}{\frac{U^\prime}{U}}+{\frac{1}{2}}{\frac{f^\prime}{f}}\right)(v_2^\prime+v_3^\prime)+{\frac{ p^2}{2U}}e^{-2v_1}-{\frac{p^2}{4U}}e^{2(v_2-v_1-v_3)}\n
  & &-{\frac{p^2}{4U}}e^{2(v_3-v_1-v_2)}+ {\frac{1}{2}}\left({\frac{f^{\prime\prime}(r)}{f(r)}} + {\frac{1}{2}}\left\lbrack{\frac{U^\prime(r)f^\prime(r)}{U(r)f(r)}}-{\frac{(f^\prime(r))^2}{(f(r))^2}}\right\rbrack\right)+{\frac{6}{U}}+{\frac{3 p^2}{8 U}}e^{-2{(v_3+v_1)}}w(r)^2 - {\frac{p^2}{8}}e^{-2 v_2}w^{\prime 2}\n
  & & +{\frac{1}{4}}\phi^{\prime 2}-{\frac{a(r)^2}{4 U(r)f(r)}}\phi^\dag \phi + {\frac{1}{4U}}e^{-2 v_2}w(r)^2 (\phi^\dag \phi)
  + {\frac{m^2}{4U}}\phi^\dag\phi              + {\frac{a^{\prime 2}}{8f(r)}},\n
0 &=&{\frac{f^\prime}{2f}}(v_1^\prime + v_2^\prime +v_3^\prime)+(v_1^\prime v_2^\prime + v_2^\prime v_3^\prime +v_3^\prime v_1^\prime) + {\frac{p^2}{U}}e^{-2 v_1}Sin\,h^2(v_2-v_3) + {\frac{3}{8}}e^{-2 v_2}w^\prime - 
{\frac{p^2}{8U}} e^{-2(v_1+v_3)} w^2\n
  & &+{\frac{1}{4}}\partial_r\phi^\dag \partial_r \phi+{\frac{1}{4}}{\frac{a(r)^2}{f(r)U(r)}}\phi^\dag\phi
                                            -{\frac{1}{4U}}e^{-2 v_2}w^2 \phi^\dag\phi -
{\frac{M^2}{4U}}\phi^\dag \phi-{\frac{3}{8}}{\frac{a^{\prime 2}}{f(r)}},\n
0 &=& \left\lbrack v_1^{\prime\prime} +v_2^{\prime\prime} +v_3^{\prime\prime} + v_1^{\prime^2}+v_2^{\prime^2}+v_3^{\prime^2}\right\rbrack +{\frac{U^\prime}{2U}}\left\lbrack v_1^\prime +v_2^\prime + v_3^\prime \right\rbrack + v_1^\prime v_2^\prime + v_2^\prime v_3^\prime +v_1^\prime v_3^\prime - {\frac{1}{2U}}p^2 e^{-2v_1}\n
  & & +{\frac{p^2}{2U}}e^{2(v_2-v_3 -v_1)}+{\frac{p^2}{2U}}e^{2(v_3-v_2 -v_1)}-{\frac{1}{8}}e^{-2v_2}{(\partial_r w(r))^2}-{\frac{p^2}{8U}}
e^{-2(v_1 + v_3)} w^2 - {\frac{1}{4fU}}a^2\phi^\dag\phi -{\frac{1}{4}}{\partial_r \phi^\dag\partial_r\phi}\n
  & & - {\frac{1}{4U}}e^{-2v_2}w^2 \phi^\dag\phi-{\frac{1}{4U}}m^2\phi^\dag\phi {\frac{3}{8}}{\frac{a^{\prime 2}}{f}}.
\la{eom1} 
\eer

\subsection{  U(1) gauge field solution}
\vskip1mm
Here we consider the equation of motion of U(1) gauge field, find a solution of the field in terms of the potentials.  Since we obtain the expression of the potential at perturbative level, so this actually makes it feasible to obtain the expression of U(1) gauge field near critical point.
The equation of motion for U(1) gauge field a(r) is
\be
a^{\prime \prime}(r) +\left\lbrace v_1^\prime + v_2^\prime + v_3^\prime +{\frac{1}{2}}\left({\frac{U^\prime}{U}}\,- {\frac{f^\prime}{f}} \right) \right\rbrace a^\prime (r) - e^{-v_1 (r) - v_2 (r) -v_3 (r)} \sqrt{{\frac{f}{U}}} w(r) {w^\prime} (r) - \left\lbrace{\frac{\phi^\dag (r) \phi(r)}{U(r)}}\right\rbrace\,a(r)=0.
\la{aeqn}
\ee
In our fixed point ansatz, we choose, around the fixed point $w(r)=0$. In the next section, we have established that near critical point $w(r)^2$ term can be ignored.  So expressing the derivatives of the metric in terms of the potentials (\ref{fieldderivative}) and (\ref{gravityequation}), we can express (\ref{aeqn}) as 
\ber
& &{\frac{d}{dr}}\left\lbrace{\frac{-2a(r)\phi^\dag (r) \phi(r)}{f(r)}} M\left(\phi^\dag (r) \phi(r) \right)\right\rbrace +{\frac{W\left(\phi^\dag (r) \phi(r)\right)}{2(d-1)}}\left\lbrace{\frac{-2a(r)\phi^\dag (r) \phi(r)M\left(\phi^\dag (r) \phi(r) \right)}{f(r)}}\right\rbrace\n
&-& \left\lbrace{\frac{\phi^\dag (r) \phi(r)}{U(r)}}\right\rbrace a(r) = 0.
\la{areducedeqn}
\eer

In order to solve the above equation we substitute
\be
{\frac{-2a(r)\phi^\dag (r) \phi(r)}{f(r)}} M\left(\phi^\dag (r) \phi(r)\right) = X.
\la{x}
\ee
Since at fixed point $M\left(\phi^\dag (r) \phi(r)\right)$ vanishes so we can write 
$$M\left(\phi^\dag (r) \phi(r)\right)= (\phi^\dag\phi - \alpha) + O\left((\phi^\dag\phi -\alpha)^2 \right)$$
Near critical point $M\left(\phi^\dag (r) \phi(r)\right)$ can be approximated by quadratic part only upto an overall constant.  
Considering a solution $U(r) = f(r)$, we can solve the above near critical point, by integrating the equation with the above prescribed value of M, as follows 
\ber
{\frac{d X}{dr}}
&+& {\frac{W\left(\phi^\dag (r) \phi(r)\right)}{2(d-1)}}X
- {\frac{X}{2 M\left(\phi^\dag (r) \phi(r)\right) }} = 0\n
ln X
&=& \int dr^\prime \left\lbrace {\frac{W\left(\phi^\dag (r^\prime) \phi(r^\prime)\right)}{2(d-1)}}
- {\frac{1}{2 M\left(\phi^\dag (r^\prime) \phi(r^\prime)\right) }}\right\rbrace\n
a(r)
&=& \left\lbrack {\frac{-2 \phi^\dag (r) \phi(r)}{f(r)}} M\left(\phi^\dag (r) \phi(r)\right) \right\rbrack ^{-1} \n
& & .Exp \left\lbrack \int dr^\prime \left\lbrace {\frac{W\left(\phi^\dag (r^\prime) \phi(r^\prime)\right)}{2(d-1)}}
- {.\frac{1}{2 M\left(\phi^\dag (r^\prime) \phi(r^\prime)\right) }}\right\rbrace\right\rbrack.
\la{areduced}
\eer
In the above we have chosen $U(r) = f(r)$, as given in (\ref{metric}) 
Near critical point we can substitute the expression of $\phi(r),M\left(\phi^\dag (r) \phi(r)\right), W\left(\phi^\dag (r) \phi(r)\right)$ 
to obtain the expression of $a(r)$.

\subsection{Gravity equations}

Here we write the gravity equations in terms of potentials. Here we use (\ref{fieldderivative},\ref{gravityequation},\ref{metricrelation}) to rewrite the equations. Also in writing these equations we replace 
\ber
\left\lbrack{\frac {d w\left(\phi^\dag (r) \phi(r)\right)}{dr}}\right\rbrack^2 
&=&\left\lbrack {\frac{d}{d\phi^\dag (r)}}\left\lbrace e^{-2 v_2 (r)} (D_i \phi (r))^\dag (D_i \phi (r))\right\rbrace \right\rbrack \left\lbrack {\frac{d}{d\phi (r)}}\left\lbrace e^{-2 v_2 (r)} (D_i \phi (r))^\dag (D_i \phi (r))\right\rbrace \right\rbrack  \n 
&=& \left\lbrace e^{-2 v_2(r)} \phi^\dag (r) \phi(r) w(r)M\left(\phi^\dag (r) \phi(r) \right) w(r) \right\rbrace^2\n
\left\lbrack{\frac {d a\left(\phi^\dag (r) \phi(r)\right)}{dr}}\right\rbrack^2 
&=&\left\lbrace {\frac{ 2 a(r)}{f(r)}}\phi^\dag (r) \phi(r) M\left(\phi^\dag (r) \phi(r) \right) \right\rbrace^2 .  
\la{wderivative}
\eer

Following (\ref{metric}), we write the equation of $g_{22} + g_{33}+g_{23}/Sin[p x_1]Cos[p x_1]$

\ber
& &-2U(r)\left\lbrace {\frac{2}{d-1}}{\frac{d}{d r}}\left\lbrack W\left(\phi^\dag(r) \phi(r)\right)\right\rbrack + {\frac{3}{(d-1)^2}}\left\lbrack W\left(\phi^\dag(r) \phi(r)\right) \right\rbrack^2 \right\rbrace + {{p^2}e^{-2 v_1}}\n 
&+& {\frac{p^2}{2}}{e^{2( v_2-v_1 -v_3)}} - {\frac{3 p^2}{2}}{e^{2( v_3-v_1 -v_3)}} -{\frac{1}{d-1}} U(r)\left\lbrace {\frac{d^2 W\left(\phi^\dag(r) \phi(r)\right)}{d {\phi^\dag}(r) d\phi(r)}}\right\rbrace -12\n
& & -{\frac{3}{4}}{{p^2}e^{-2 v_1}}{[w(r)]}^2 {e^{-2 v_3}} + U(r) e^{-2 v_2(r)}\left\lbrace e^{-2 v_2(r)} \phi^\dag (r) \phi(r) w(r)M\left(\phi^\dag (r) \phi(r) \right)\right\rbrace^2 \n
& & +{\frac{1}{2}}\phi^\dag (r) \phi(r)e^{-2 v_2(r)} w(r)^2 + {\frac{U(r)}{2}}{\frac{\partial  W\left(\phi^\dag (r) \phi(r) \right)}{\partial \phi^\dag (r)}}{\frac{\partial  W\left(\phi^\dag (r) \phi(r) \right)}{\partial \phi (r)}} -{\frac{a(r)^2}{2 f(r)}}\left(\phi^\dag (r) \phi(r) \right)\n
&-& {\frac{1}{4}}\left\lbrace {\frac{ 2 a(r)}{f(r)}}\phi^\dag (r) \phi(r) M\left(\phi^\dag (r) \phi(r) \right) \right\rbrace^2  
+ {\frac{1}{2}}m^2 \phi^\dag (r) \phi(r)\n
&=& 0.
\la{gravitysuperpotential1}
\eer

Following (\ref{metric}), we write the equation of $g_{22} + g_{33}-g_{23}/Sin[p x_1]Cos[p x_1]$

\ber
& &-2U(r)\left\lbrace {\frac{2}{d-1}}{\frac{d}{d r}}\left\lbrack W\left(\phi^\dag(r) \phi(r)\right)\right\rbrack + {\frac{3}{(d-1)^2}}\left\lbrack W\left(\phi^\dag(r) \phi(r)\right) \right\rbrack^2\right\rbrace + {{p^2}e^{-2 v_1}}\n 
&-& {\frac{3 p^2}{2}}{e^{2( v_2-v_1 -v_3)}} + {\frac{ p^2}{2}}{e^{2( v_3-v_1 -v_2)}} -{\frac{1}{d-1}} U(r)\left\lbrace {\frac{d^2  W\left(\phi^\dag(r) \phi(r)\right)}{d {\phi^\dag}(r) d\phi(r)}}\right\rbrace -12\n
& &  - {\frac{3 U(r)}{4}}\left\lbrace 2 e^{-2 v_2(r)} \phi^\dag (r) \phi(r) w(r)M\left(\phi^\dag (r) \phi(r) \right)\right\rbrace^2 \n
& & + {\frac{1}{4}} p^2 e^{-2 v_1(r)-2 v_3(r)} w(r)^2  
+{\frac{U(r)}{2}}{\frac{\partial  W\left(\phi^\dag (r) \phi(r) \right)}{\partial \phi^\dag (r)}}{\frac{\partial  W\left(\phi^\dag (r) \phi(r) \right)}{\partial \phi (r)}}-{\frac{a(r)^2}{2 f(r)}}\left(\phi^\dag (r) \phi(r) \right)\n
&-& {\frac{1}{4}}\left\lbrace {\frac{ 2 a(r)}{f(r)}}\phi^\dag (r) \phi(r) M\left(\phi^\dag (r) \phi(r) \right) \right\rbrace^2  -{\frac{1}{2}}\phi^\dag (r) \phi(r)e^{-2 v_2(r)} w(r)^2
+ {\frac{1}{2}}m^2 \phi^\dag (r) \phi(r)\n
&=& 0.
\la{gravitysuperpotential2}
\eer
The equation of motion of $g_{x^1 x^1}$ component of metric is

\ber
& &-2U(r)\left\lbrace {\frac{2}{d-1}}{\frac{d}{d r}}\left\lbrack W\left(\phi^\dag(r) \phi(r)\right)\right\rbrack + {\frac{3}{(d-1)^2}}\left\lbrack W\left(\phi^\dag(r) \phi(r)\right)\right\rbrack^2\right\rbrace - {{p^2}e^{-2 v_1}}\n 
&+& {\frac{ p^2}{2}}{e^{2( v_2-v_1 -v_3)}} + {\frac{ p^2}{2}}{e^{2( v_3-v_1 -v_2)}} 
+{\frac{1}{d-1}} U(r)\left\lbrace {\frac{d W\left(\phi^\dag(r) \phi(r)\right)}{d {\phi^\dag}(r) d\phi(r)}}\right\rbrace -12\n
& &  + {\frac{ U(r)}{4}} \left\lbrace e^{-2 v_2(r)} \phi^\dag (r) \phi(r) w(r)M\left(\phi^\dag (r) \phi(r) \right)\right\rbrace^2 \n
& & - {\frac{3}{4}} p^2 e^{-2 v_1(r)-2 v_3(r)} w(r)^2 - {\frac{U(r)}{2}}{\frac{\partial  W\left(\phi^\dag (r) \phi(r) \right)}{\partial \phi^\dag (r)}}{\frac{\partial  W\left(\phi^\dag (r) \phi(r) \right)}{\partial \phi (r)}} + {\frac{a(r)^2}{2 f(r)}}\left(\phi^\dag (r) \phi(r) \right)\n
&-& {\frac{1}{4}}\left\lbrace {\frac{ 2 a(r)}{f(r)}}\phi^\dag (r) \phi(r) M\left(\phi^\dag (r) \phi(r) \right) \right\rbrace^2  
-{\frac{1}{2}}\phi^\dag (r) \phi(r)e^{-2 v_2(r)} w(r)^2 + {\frac{1}{2}}m^2 \phi^\dag (r) \phi(r)\n 
&=& 0
\la{superpotential3}
\eer

Finally to find an estimate of $g_{tt}= - f(r)$ at critical point with $f(r) = U(r)$,  we proceed with the  equation of motion of $g_{tt}$ component, in terms of the potentials, ignoring $w(r)$ square terms,
\ber
& &-\left\lbrace {\frac{3}{d-1}}{\frac{d}{d r}}\left\lbrack W\left(\phi^\dag(r) \phi(r)\right)\right\rbrack + {\frac{3}{4(d-1)^2}}\left\lbrack W\left(\phi^\dag(r) \phi(r)\right) \right\rbrack^2\right\rbrace  -6\n
 &-& {\frac{1}{4}}{\frac{\partial  W\left(\phi^\dag (r) \phi(r) \right)}{\partial \phi^\dag (r)}}{\frac{\partial  W\left(\phi^\dag (r) \phi(r) \right)}{\partial \phi (r)}}
\n
&-&{\frac{a(r)^2}{4 f^2 (r)}}\left(\phi^\dag (r) \phi(r) \right)\n
&=& {\frac{3}{8 f(r)}}\left\lbrace {\frac{ 2 a(r)}{f(r)}}\phi^\dag (r) \phi(r) M\left(\phi^\dag (r) \phi(r) \right) \right\rbrace^2  
+ {\frac{1}{4 f(r)}}m^2 \phi^\dag (r) \phi(r).  
\la{gravitysuperpot}
\eer

Subtracting  (\ref{gravitysuperpotential2}) from (\ref{gravitysuperpotential1}) gives

\ber
& & 2 p^2 \left\lbrack {e^{2( v_2-v_1 -v_3)}} 
- {e^{2( v_3-v_1 -v_2)}} \right\rbrack
 + U(r)\left\lbrace 2 e^{-2 v_2(r)} \phi^\dag (r) \phi(r) w(r)M\left(\phi^\dag (r) \phi(r) \right)\right\rbrace^2\rbrack\n 
-  p^2 e^{-2 v_1(r)-2 v_3(r)} w(r)^2
&+& {\frac{1}{2}}\phi^\dag (r) \phi(r)e^{-2 v_2(r)} w(r)^2 
= 0.
\la{firstdifferenceeqn}
\eer
 Now in order to understand the above relation, first we expand $w(r)$ around its fixed point value $w(\phi_*)$.

For any r, we can write
\ber
w(r) 
&=& w(\phi_*) +\left\lbrace {\frac{\partial w(r)}{\partial \phi(r)}}{\frac{\partial \phi(r)}{\partial r}}{|_{\phi(r)=\phi_*}} (\phi(r) - \phi_*) + {\frac{\partial w(r)}{\partial \phi^\dag (r)}}{\frac{\partial \phi^\dag (r)}{\partial r}}{|_{\phi(r)=\phi_*}} (\phi^\dag (r) - \phi_*^*)\right\rbrace\n
&+&.......
\la{wexpansion}
\eer

Now we choose fixed point 
\be
w(\phi_*) = 0
\la{wcritical}
\ee
 Then from (\ref{fieldderivative}) we also have first derivative of $w(r)$ vanishes at critical  point.  So $w(r)^2$ terms are actually of the order $\lbrack \left( \phi(r)^\dag \phi(r) - \alpha \right)^4\rbrack $

Since we are studying fixed point at quadratic order, so we can actually ignore the ${w(r)^2}$ term in (\ref{firstdifferenceeqn})
and write the eqn as
\be
p^2 \left\lbrack {e^{2( v_2-v_1 -v_3)}} 
- {e^{2( v_3-v_1 -v_2)}} \right\rbrack = 0 .
\la{relation}
\ee

So finally we are having the relation, near fixed point
\be
v_2 (r) = v_3 (r).
\la{vrelation}
\ee

Next we substract (\ref{superpotential3}) from (\ref{gravitysuperpotential2}) to obtain
\ber
-& &{\frac{U(r)}{2}}{\frac{\partial  W\left(\phi^\dag (r) \phi(r) \right)}{\partial \phi^\dag (r)}}{\frac{\partial  W\left(\phi^\dag (r) \phi(r) \right)}{\partial \phi (r)}} 
+ {\frac{a(r)^2}{2 f(r)}}\left(\phi^\dag (r) \phi(r) \right) 
= 0 \n
\left\lbrack {\frac{a(r)}{f(r)}}\right\rbrack^2
 &=& {\frac{1}{\left(\phi^\dag (r) \phi(r) \right) }} {\frac{\partial  W\left(\phi^\dag (r) \phi(r) \right)}{\partial \phi^\dag (r)}}{\frac{\partial  W\left(\phi^\dag (r) \phi(r) \right)}{\partial \phi (r)}}. 
\la{abyf}
\eer
in the above  we use  $U(r)= f(r)$ to obtain ${\frac{a(r)}{f(r)}}$ in terms of W.

\ber
{\frac{1}{ f(r)}}
&=& \left\lbrack {\frac{3}{8 }}\left\lbrace {\frac{ 2 a(r)}{f(r)}}\phi^\dag (r) \phi(r) M\left(\phi^\dag (r) \phi(r) \right) \right\rbrack^2  
+ {\frac{1}{4 }} m^2 \phi^\dag (r) \phi(r)\right\rbrack^{-1}\n
& &\lbrack -\left\lbrace {\frac{3}{d-1}}{\frac{d}{d r}}\left\lbrack W\left(\phi^\dag (r) \phi (r)\right)\right\rbrack + 
{\frac{3}{4(d-1)^2}}\left\lbrack W\left(\phi^\dag(r) \phi(r)\right) \right\rbrack^2\right\rbrace  -6\n
 &-& {\frac{1}{4}}{\frac{\partial W\left(\phi^\dag (r) \phi(r) \right)}{\partial \phi^\dag (r)}}{\frac{\partial W\left(\phi^\dag (r) 
\phi(r) \right)}{\partial \phi (r)}}
\n
&-&{\frac{a(r)^2}{4 f^2 (r)}}\left(\phi^\dag (r) \phi(r) \right)\rbrack^{-1} 
\la{gtt}
\eer

Here once again we ignore the quartic part of $w(r)$ which add higher order correction.

{}

\begin{thebibliography}{99}

\bibitem{kiritsis} 
  Elias Kiritsis, Wenliang Li, Francesco Nitti,
  ``Holographic RG flow and quantum effective action,''
  Fortsch.\ Phys. {\bf 62},(2014)389-454
  doi:10.1002/prop.201400007
  [arXiv:1401.0888 [hep-th]].
\bibitem{exotic} 
  Elias Kiritsis, Francesco Nitti, Leandro Silva Pimenta,
  ``Exotic RG flow from holography,''
  Fortsch.\ Phys. {\bf 65},(2017)no.2, 1600120
  doi:10.1002/prop.201600120
  [arXiv:1611.05493 [hep-th]].

\bibitem{kikuchi} 
  Ken Kikuchi, Tadakatsu Sakai,
  ``AdS/CFT and local renormalization group with gauge field,''
  PTEP 2016 {\bf 2016},no.3, 033B02
  doi:10.1093/ptep/ptw010
  [arXiv:1511.00403 [hep-th]].

\bibitem{dyonic} 
  E.J. Lindgren, Ioannis Papadimitriou, Anastasios Taliotis, Joris Vanhoof,
  ``Holographic Hall conductivities from dyonic backgrounds,''
  JHEP 1507{\bf 094}2015
  doi: 10.1007/JHEP07(2015)094
  [arXiv:1505.04131 [hep-th]].

\bibitem{lectholo2} 
  Jan de Boer,
  ``The Holographic Renormalization Group,''
 Fortsch.\ Phys. {\bf 49},(2001)339-358
  doi: 10.1002/1521-3978(200105)49:4/6<339::AID-PROP339>3.0.CO;2-A 
  [arXiv:0101026 [hep-th]].

\bibitem{verlinde} 
  J. de Boer, E. Verlinde, H. Verlinde,
  ``On the Holographic Renormalization Group,''
 JHEP 0008 {\bf 003},(2000)
  doi: 10.1088/1126-6708/2000/08/003 
  [arXiv:9912012 [hep-th]].

\bibitem{papa} 
  Ioannis Papadimitriou, Kostas Skenderis,
  ``Correlation Functions in Holographic RG Flows,''
 JHEP 0410 {\bf 075},(2004)
  doi:  10.1088/1126-6708/2004/10/075 
  [arXiv:0407071[hep-th]].

\bibitem{Gouteraux} 
  B. Goutéraux, E. Kiritsis,
  ``Quantum critical lines in holographic phases with (un)broken symmetry,''
 JHEP 1304 {\bf 053},(2013)
  doi:  10.1007/JHEP04(2013)053  
  [arXiv:1212.2625[hep-th]].

\bibitem{nakayama} 
  Yu Nakayama,
  ``Vector Beta function,''
 Int.\ J .\ Mod .\ Phys .\ A. {\bf 28},(2013)1350166
  doi:   10.1142/S0217751X13501662  
  [arXiv:1310.0574[hep-th]].

\bibitem{bourdier} 
  Jun Bourdier, Elias Kiritsis,
  ``Holographic RG flows and nearly-marginal operators,''
 Class.\ Quant .\ Grav . {\bf 31},(2014)035011
  doi:  10.1088/0264-9381/31/3/035011  
  [arXiv:1310.0858[hep-th]].

\bibitem{ac} 
 Thomas Faulkner, Hong Liu, Mukund Rangamani,
  ``Integrating out geometry: Holographic Wilsonian RG and the membrane paradigm,''
 JHEP 1108  {\bf 051},(2011)
  doi:  10.1007/JHEP08(2011)051  
  [arXiv:1010.4036[hep-th]].

\bibitem{local} 
D.Z. Freedman, S.S. Gubser, K. Pilch, N.P. Warner ,
  ``Renormalization Group Flows from Holography--Supersymmetry and a c-Theorem,''
 Adv.\ Theor.\ Math.\ Phys.  {\bf 3},(1999) 363-417
  doi: 10.4310/ATMP.1999.v3.n2.a7   
  [arXiv:9904017[hep-th]].

\bibitem{behr} 
Idse Heemskerk, Joseph Polchinski,
  ``Holographic and Wilsonian Renormalization Groups,''
 JHEP 1106  {\bf 031},(2011) 
  doi: 10.1007/JHEP06(2011)031   
  [arXiv:1010.1264[hep-th]].

\bibitem{papa} 
Ioannis Papadimitriou,
  ``Holographic Renormalization of general dilaton-axion gravity,''
 JHEP 1108  {\bf 119},(2011) 
  doi: 10.1007/JHEP08(2011)119  
  [arXiv:1106.4826[hep-th]].

\bibitem{niar} 
Elias Kiritsis, Vasilis Niarchos,
  ``The holographic quantum effective potential at finite temperature and density,''
 JHEP 1208  {\bf 164},(2012) 
  doi: 10.1007/JHEP08(2012)164
  [arXiv:1205.6205[hep-th]].

\bibitem{curtright} 
Thomas L Curtright, Xiang Jin, Cosmas K Zachos,
  ``RG flows, cycles, and c-theorem folklore,''
 Phys.\ Rev.\ Lett.  {\bf 108},(2012)131601 
  doi: 10.1103/PhysRevLett.108.131601 
  [arXiv:1111.2649[hep-th]].

\bibitem{lee} 
Sung-Sik Lee,
  ``Quantum Renormalization Group and Holography,''
 JHEP 01 (2014) {\bf 076},
  doi: 10.1007/JHEP01(2014)076
  [arXiv:1305.3908[hep-th]].


\bibitem{notes1} 
  Kostas Skenderis,
  ``Lecture Notes on Holographic Renormalization,''
 Class.\ Quant .\ Grav . {\bf 19},(2002)5849-5876
  doi: 10.1088/0264-9381/19/22/306   
  [arXiv:0209067[hep-th]].

\bibitem{notes2} 
Vijay Balasubramanian, Per Kraus,
  ``Spacetime and the Holographic Renormalization Group,''
 Phys.\ Rev.\ Lett.  {\bf 83},(1999)3605-3608 
  doi: 10.1103/PhysRevLett.83.3605
  [arXiv:9903190[hep-th]].

\bibitem{bianchi} 
Massimo Bianchi, Daniel Z. Freedman, Kostas Skenderis,
  ``How to go with an RG Flow,''
 JHEP 0108 (2014) {\bf 076},
  doi: 10.1007/JHEP01(2001)041
  [arXiv:0105276[hep-th]].


\bibitem{Hartnoll:2009sz} 
  S.~A.~Hartnoll,
  ``Lectures on holographic methods for condensed matter physics,''
  Class.\ Quant.\ Grav.\  {\bf 26}, 224002 (2009)
  doi:10.1088/0264-9381/26/22/224002
  [arXiv:0903.3246 [hep-th]].
  
\bibitem{Herzog:2009xv} 
  C.~P.~Herzog,
  ``Lectures on Holographic Superfluidity and Superconductivity,''
  J.\ Phys.\ A {\bf 42}, 343001 (2009)
  doi:10.1088/1751-8113/42/34/343001
  [arXiv:0904.1975 [hep-th]].
  
\bibitem{Horowitz:2010gk} 
  G.~T.~Horowitz,
  ``Introduction to Holographic Superconductors,''
  Lect.\ Notes Phys.\  {\bf 828}, 313 (2011)
  doi:10.1007/978-3-642-04864-7-10
  [arXiv:1002.1722 [hep-th]].
  
\bibitem{Hartnoll:2011fn} 
  S.~A.~Hartnoll,
  ``Horizons, holography and condensed matter,''
  arXiv:1106.4324 [hep-th].
  
\bibitem{Sachdev:2011wg} 
  S.~Sachdev,
  ``What can gauge-gravity duality teach us about condensed matter physics?,''
  Ann.\ Rev.\ Condensed Matter Phys.\  {\bf 3}, 9 (2012)
  doi:10.1146/annurev-conmatphys-020911-125141
  [arXiv:1108.1197 [cond-mat.str-el]].
  
\bibitem{Green:2013fqa} 
  A.~G.~Green,
  ``An Introduction to Gauge Gravity Duality and Its Application in Condensed Matter,''
  Contemp.\ Phys.\  {\bf 54}, no. 1, 33 (2013)
  doi:10.1080/00107514.2013.779477
  [arXiv:1304.5908 [cond-mat.str-el]].

\bibitem{buildingads} 
Sean A. Hartnoll, Christopher P. Herzog, Gary T. Horowitz,
  ``Building an AdS/CFT superconductor,''
 Phys.\ Rev.\ Lett.  {\bf 101},(2008)031601 
  doi: 10.1103/PhysRevLett.101.031601 
  [arXiv:0803.3295[hep-th]].

\bibitem{hertog} 
Thomas Hertog,
  ``Towards a Novel no-hair Theorem for Black Holes,''
 Phys.\ Rev.\ D.  {\bf 74},(2006) 084008 
  doi: 10.1103/PhysRevD.74.084008 
  [arXiv:0608075[gr-qc]].

\bibitem{gubser} 
Steven S. Gubser,
  ``Breaking an Abelian gauge symmetry near a black hole horizon,''
 Phys.\ Rev.\ D.  {\bf 78},(2008) 065034  
  doi:10.1103/PhysRevD.78.065034  
  [arXiv:0801.2977[ hep-th]].

\bibitem{horowitz} 
Gary T. Horowitz, Matthew M. Roberts,
 ``Holographic Superconductors with Various Condensates,''
 Phys.\ Rev.\ D.  {\bf 78},(2008) 126008 
  doi:10.1103/PhysRevD.78.126008 
  [arXiv:0810.1077[ hep-th]].

\bibitem{roberts} 
Gary T. Horowitz, Matthew M. Roberts,
  ``Zero Temperature Limit of Holographic Superconductors,''
 JHEP 0911 (2009) {\bf 015},
  doi:  10.1088/1126-6708/2009/11/015
  [arXiv:0908.3677[hep-th]].

\bibitem{horo} 
Thomas Faulkner, Gary T. Horowitz, John McGreevy, Matthew M. Roberts, David Vegh,
  ``Photoemission "experiments" on holographic superconductors,''
 JHEP 1003 (2010) {\bf 121},
  doi:  10.1007/JHEP03(2010)121
  [arXiv:0911.3402[hep-th]].

\bibitem{nellore} 
Steven S. Gubser, Abhinav Nellore,
 ``Ground states of holographic superconductors,''
 Phys.\ Rev.\ D.  {\bf 80},(2009) 105007
  doi:10.1103/PhysRevD.80.105007 
  [arXiv:0908.1972[ hep-th]].

\bibitem{denef} 
Frederik Denef, Sean A. Hartnoll, Subir Sachdev,
 ``Quantum oscillations and black hole ringing,''
 Phys.\ Rev.\ D.  {\bf 80},(2009) 126016
  doi:10.1103/PhysRevD.80.126016
  [arXiv:0908.1788[ hep-th]].


\bibitem{pufu} 
Steven S. Gubser, Christopher P. Herzog, Silviu S. Pufu, Tiberiu Tesileanu,
  ``Superconductors from Superstrings,''
 Phys.\ Rev.\ Lett.  {\bf 103},(2009)141601 
  doi: 10.1103/PhysRevLett.103.141601
  [arXiv:0907.3510[hep-th]].

\bibitem{metal2insulator} 
Aristomenis Donos, Sean A. Hartnoll,
  ``Metal-insulator transition in holography,''
 Nature Phys.  {\bf 9},(2013)649-655 
  doi: 10.1038/nphys2701
  [arXiv:1212.2998[hep-th]].

\bibitem{metalinsulator} 
Aristomenis Donos, Blaise Goutéraux, Elias Kiritsis,
  ``Holographic Metals and Insulators with Helical Symmetry,''
 JHEP 1409 (2014) {\bf 038},
  doi: 10.1007/JHEP09(2014)038 
  [arXiv:1406.6351[hep-th]].

\bibitem{gaun} 
Aristomenis Donos, Jerome P. Gauntlett,
  ``Holographic helical superconductors,''
 JHEP 1112 (2011) {\bf 091},
  doi: 10.1007/JHEP12(2011)091 
  [arXiv:1109.3866[hep-th]].

\bibitem{t} 
Aristomenis Donos, Jerome P. Gauntlett,
  ``Helical superconducting black holes,''
 Phys.\ Rev.\ Lett.  {\bf 108},(2012)211601 
  doi:10.1103/PhysRevLett.108.211601 
  [arXiv:1203.0533[hep-th]]. 

\bibitem{lett} 
Aristomenis Donos, Jerome P. Gauntlett,
 ``Black holes dual to helical current phases,''
 Phys.\ Rev.\ D.  {\bf 86},(2012)064010 
  doi:10.1103/PhysRevD.86.064010
  [arXiv:1204.1734[ hep-th]].

\bibitem{iqbal} 
Nabil Iqbal, Hong Liu, Márk Mezei, Qimiao Si,
 ``Quantum phase transitions in holographic models of magnetism and superconductors,''
 Phys.\ Rev.\ D.  {\bf 82},(2010)045002 
  doi:10.1103/PhysRevD.82.045002
  [arXiv:1003.0010[ hep-th]].

\bibitem{oupar} 
  Subir Mukhopadhyay, Chandrima Paul,
  ``Phases in holographic helical superconductor,''
 Int.\ J .\ Mod .\ Phys .\ A. {\bf 32},(2017)no.13, 1750064
  doi:  10.1142/S0217751X17500646.

\bibitem{fermi} 
Thomas Faulkner, Hong Liu, John McGreevy, David Vegh,
 ``Emergent quantum criticality, Fermi surfaces, and $AdS_2$,''
 Phys.\ Rev.\ D.  {\bf 83},(2011)125002 
  doi:10.1103/PhysRevD.83.125002
  [arXiv:0907.2694[ hep-th]].

\end{thebibliography}
\end{document}